# Programmable DNA-Mediated Multitasking Processor

*Jian-Jun SHU**,†, Qi-Wen WANG†, Kian-Yan YONG†, Fangwei SHAO‡, and Kee Jin LEE†*

†School of Mechanical & Aerospace Engineering, Nanyang Technological University, 50 Nanyang Avenue, Singapore 639798

‡School of Physical & Mathematical Sciences, Nanyang Technological University, 21 Nanyang Link, Singapore 637371

KEYWORDS: DNA; processor; material; programmable biochemical operator.

ABSTRACT: Because of DNA appealing features as perfect material, including minuscule size, defined structural repeat and rigidity, programmable DNA-mediated processing is a promising computing paradigm, which employs DNAs as information storing and processing substrates to tackle the computational problems. The massive parallelism of DNA hybridization exhibits transcendent potential to improve multitasking capabilities and yield a tremendous speed-up over the conventional electronic processors with stepwise signal cascade. As an example of multitasking capability, we present an *in vitro* programmable DNA-mediated optimal route planning processor as a functional unit embedded in contemporary navigation systems. The novel programmable DNA-mediated processor has several advantages over the existing silicon-mediated methods, such as conducting massive data storage and simultaneous processing *via* much fewer materials than conventional silicon devices.

## ■ INTRODUCTION

For a conventional silicon-mediated digital computer, the growing demands for computational ability, processing speed and parallelism, require the size of individual transistor elements to be significantly reduced and hence to allow additional elements to be packed onto the same chip. The increasing packing density has led to many essential problems, including power consumption and heat dissipation. More alarmingly, the entire semiconductor industry is quickly approaching the physical constraints as predicted by Moore's law.[1] In principle, any device endowed with three fundamental functions – processing, storing and displaying information – can be regarded as a computer. Therefore, researchers from various disciplines are engaged in exploring alternatives to silicon-mediated digital computer.[2-4] Among various intriguing approaches, DNA-mediated computing seems to be the feasible strategy to compete with silicon-mediated counterpart, and ultimately brings the entire field into a new era. DNA has appealing features,[5] including minuscule size, defined structural repeat and rigidity. Programmable DNA-mediated processing is a promising computing paradigm, which employs DNAs as information storing and processing substrates to tackle the computational problems. Since the demands of monolithic parallel computing ability have grown rapidly due to specific computational algorithm,[6] the massive parallelism of DNA hybridization exhibits transcendent potential to improve multitasking capabilities and yield a tremendous speed-up over the traditional electronic processors with stepwise signal cascade. The novel programmable DNA-mediated processor has several advantages over the existing silicon-mediated methods, such as conducting massive data storage[7] and simultaneous processing[8] *via* much fewer materials than conventional silicon devices. As compared with the history of evolving silicon-mediated computer, the development of DNA-mediated computer remains at relatively early stage.[9] DNA molecules have been successfully utilized to demonstrate the solutions of various problems, such as Hamiltonian path problem,[10] maximal clique problem[11] and strategic assignment problem.[12] As an example of multitasking capability, an *in vitro* programmable DNA-mediated optimal route planning processor is designed and experimentally demonstrated.

## ■ METHODS

Optimal route planning processor behaves as a functional unit embedded in contemporary GPS (Global Positioning System) navigator. By assigning any two physical locations, present location and final destination, to the processor, it automatically routes an optimal path, or a shortest path, between the selected places, based on the information pre-stored within its database. The programmable DNA-mediated optimal route planning processor is employed to perform the equivalent function by means of using DNA molecules as information storing and processing instrument. For demonstration purpose, the selected case study is an arbitrary map in GPS navigation system, which contains exactly six physical locations connected by the bi-paths with opposite directions, as shown in Figure 1. For the sake of simplicity, the key information on the map can be represented by using an abstract graph S, as shown in the right hand side of Figure 1. Each vertex $i$ and weighted edge $(i, j)$ represent an associated physical location and the connecting path from location $i$ to $j$, respectively, in the original map, where $i, j \in \{1, 2, 3, 4, 5,$ 6}.  The numbers placed on the side of the edges are the lengths (or weights) of the corresponding connecting path.  The entire case study can be converted to a mathematical problem:  Given a graph S, made up of vertices and weighted edges, what is the shortest path between any two user-specified vertices?

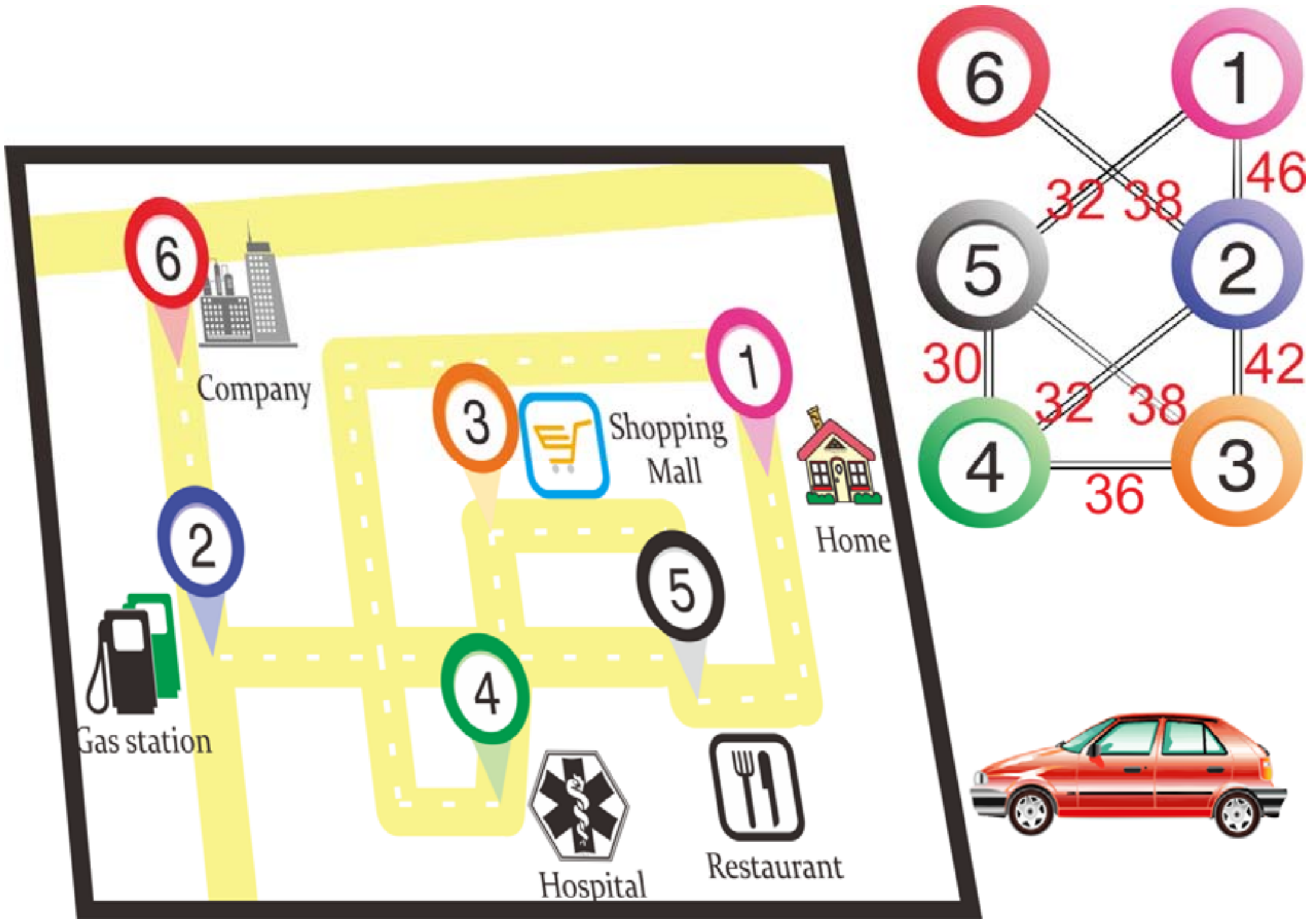

**Figure 1.** Scheme of an optimal route planning problem:  The original map with six locations (left) can be converted to an abstract graph (right) with each vertices indicated as colored circles and the length of each bi-path annotated by red digits, respectively.

The programmable DNA-mediated optimal route planning apparatus is composed of six stages as shown in Figure 2.  1).  The constituent elements, vertices and weighted edges in the graph S, are converted to different DNA sequences by *problem encoder*.  2).  Based on the information obtained from *problem encoder*, DNA solution required for subsequent processes is prepared in *DNA solution bay*. 3). In *mixing controller*, the appropriate DNA solution of vertices and weighted edges is mixed and ligased to generate a template pool of DNA duplexes that represent the possible routes from any given start to destination. In the perspective of mathematics, all non-restricted random walks are initiated within the graph S upon the completion of hybridization. 4). The obtained DNA template containing the optimal path is isolated from the oligonucleotides of incomplete hybridization, enzymes and other possible impurities. 5). The optimal path is subsequently amplified by using PCR (Polymerase Chain Reaction). The input commands to the processor are translated into the forward DNA primer representing the initial location and the reverse DNA primer representing the destination. As the efficiency of PCR is proportional to the length and copy number of the DNA templates, the DNA duplex that represents the optimal path (usually the shortest path) becomes the dominate products from *PCR amplifier*. 6). The PCR result is subjected to native polyacrylamide *gel electrophoresis* to gauge the lengths of the dominate products as the optimal (shortest) path. In this stage, the length of DNA duplex is deciphered based upon the path lengths between each pair of locations, to determine the optimal route from the given start location and destination.

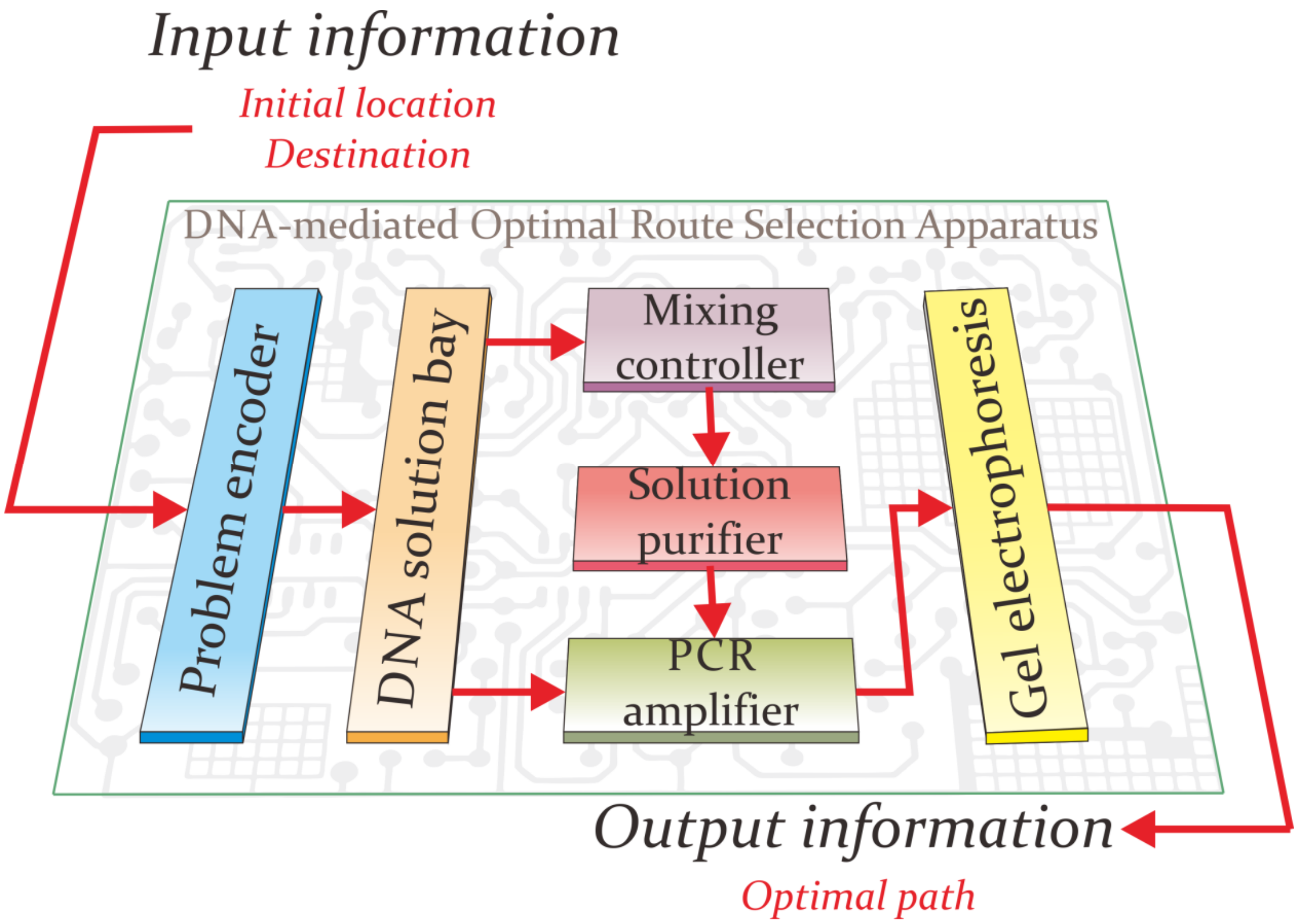

**Figure 2.** Process flow chart of programmable DNA-mediated optimal route planning processor

*Problem encoder* – For silicon-mediated computer, the very beginning stage is to digitize information in terms of binary expressions. Analogous to digital computing, the first step for DNA-mediated computing is to encode information in terms of a combination of DNA sequences – A, T, G and C.

Each vertex in the graph S is encoded into a 20-mer ssDNA (single-stranded DNA). To avoid undesired hybridization during mixing stage, the generated DNA strands have 50% CG content and melting temperature, $T_m$ = 70°C. Based on the specified rule and the sequence designing methodology,[13] the encoded DNA sequences with respect to their corresponding vertices are listed in Table 1.

**Table 1.** Vertices and the corresponding ssDNA sequences

| vertex | ssDNA sequence (5' to 3') |
|---|---|
| 1 | CCGTGTCTACAACAGAAGGA |
| 2 | CCCTCATTTGACGAGGAATG |
| 3 | TGGTTAGACGCAGAGAGTTC |
| 4 | AGGCACACGATTATGGACAG |
| 5 | CGAATTTAGCACCGCATGTG |
| 6 | GGGGTCTTCAACTATTGTCC |

As described in graph theory, the edge within the graph S is used to connect two adjacent vertices. The entire graph S is a bi-graph which means there is no specific restriction on the direction of edges. Therefore, it requires sixteen assorted DNA templates to symbolize eight corresponding edges. The template used to encode the edges is specifically designed as the dsDNA (double-stranded DNA) duplex with two overhangs (*i.e.* exposed ssDNA). As the example shown in Figure 3, each edge DNA contains a duplex portion (Segment 2) with overhangs extended from 5'- and 3'-overhangs (Segment 1 and Segment 3, respectively). Segments 1 and 3 are employed to connect from vertex *i* to vertex *j*, whereas Segment 1 is complementary to the rear 10-mer ssDNA of vertex *i* and Segment 3 is complementary to the former 10-mer ssDNA of vertex *j*, where $i, j \in \{1, 2, 3, 4, 5, 6\}$. Segment 2 is acting as the biasing parameter to count the length of the edges. The length of Segment 2 is determined by subtracting 20bp (base pair) from the length of the corresponding edges in graph S. For instance, edge (1, 5) has 32 units of length as depicted in S. Therefore, the corresponding length of Segment 2 of DNA template is 12bp.  In short, the length of edge is proportional to that of the assigned DNA template.

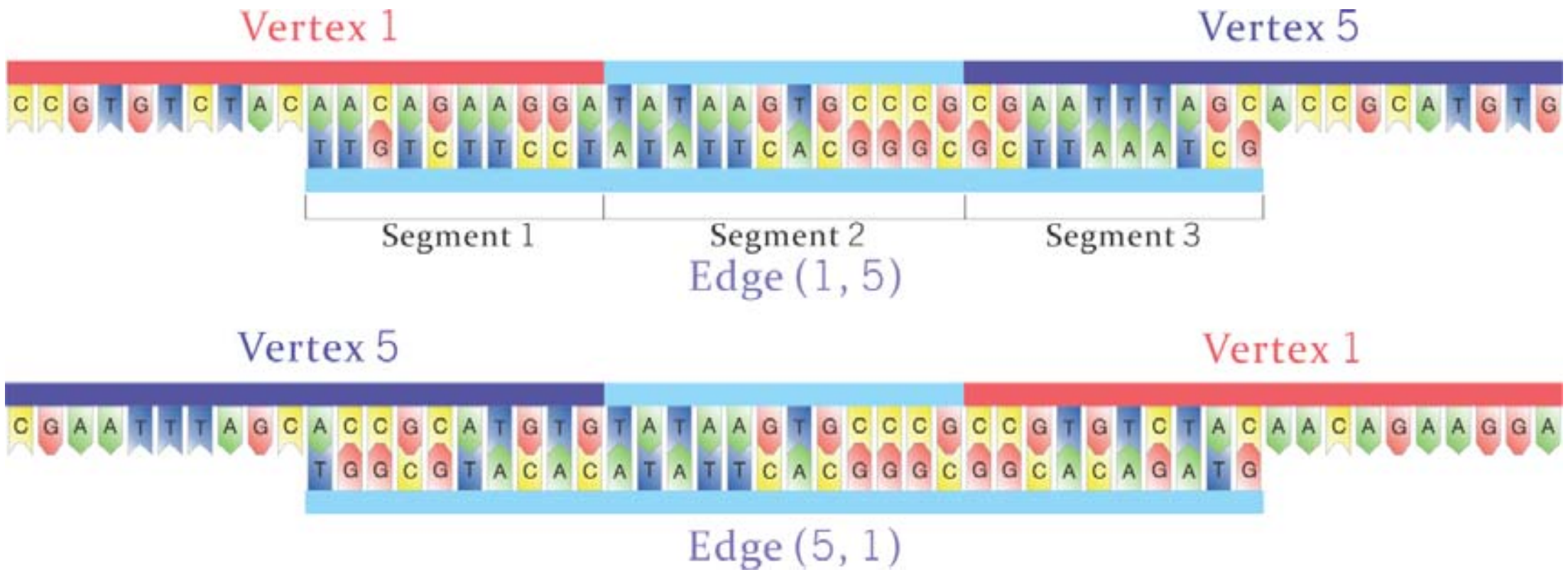

**Figure 3.** Schematic representation of edge encoding scheme.  The backbones of the vertex DNA sequences are presented in red or navy, which those of the edge DNA sequences are presented in blue.

*DNA solution bay* – In order to prevent cross-contamination, DNA solution required for the subsequent processes is prepared within an isolated and decontaminated area, called *DNA solution bay*.  The DNA strands representing vertex 1~6, sixteen edge paths and six pairs of primers for the amplification of optimal solution are listed in Tables 1, 2, and 3, respectively.  The vertex and primer DNA strands are prepared as single strands and stored as 100µM aqueous stock solution, while the edge DNA strands are annealed to duplex and prepared into 50µM stock solution.

**Table 2.** Edges and the corresponding dsDNA sequences

| **edge** | **dsDNA sequence**[#] |
|---|---|
| (1,2) | 5’-CCCAGTTAATGCAACCTTGTAGCGGT–3’<br>3’-TTGTCTTCCTGGGTCAATTACGTTGGAACATCGCCAGGGAGTAAAC-5’ |
| (2,1) | 5’-CCCAGTTAATGCAACCTTGTAGCGGT-3’<br>3’-TGCTCCTTACGGGTCAATTACGTTGGAACATCGCCAGGCACAGATG-5’ |
| (1,5) | 5’-TATAAGTGCCCG-3’<br>3’-TTGTCTTCCTATATTCACGGGCGCTTAAATCG-5’ |
| (5,1) | 5’-TATAAGTGCCCG-3’<br>3’- TGGCGTACACATATTCACGGGCGGCACAGATG-5’ |
| (2,3) | 5’-TATGCTGACCATCCTACCTTCC-3’<br>3’-TGCTCCTTACATACGACTGGTAGGATGGAAGGACCAATCTGC-5’ |
| (3,2) | 5’-TATGCTGACCATCCTACCTTCC-3’<br>3’-GTCTCTCAAGATACGACTGGTAGGATGGAAGGGGGAGTAAAC-5’ |
| (2,4) | 5’-AAAGCTCTAGGG-3’<br>3’-TGCTCCTTACTTTCGAGATCCCTCCGTGTGCT-5’ |
| (4,2) | 5’-AAAGCTCTAGGG-3’<br>3’-AATACCTGTCTTTCGAGATCCCGGGAGTAAAC-5’ |
| (2,6) | 5’-ACATGACATCTCCGCTGA-3’<br>3’-TGCTCCTTACTGTACTGTAGAGGCGACTCCCCAGAAGT-5’ |
| (6,2) | 5’-ACATGACATCTCCGCTGA-3’<br>3’-TGATAACAGGTGTACTGTAGAGGCGACTGGGAGTAAAC-5’ |
| (3,4) | 5’-ATCTGGTATTCGTCCC-3’<br>3’-GTCTCTCAAGTAGACCATAAGCAGGGTCCGTGTGCT-5’ |
| (4,3) | 5’-ATCTGGTATTCGTCCC-3’ |
| | 3'-AATACCTGTCTAGACCATAAGCAGGGACCAATCTGC-5' |
| (3,5) | 5'-CCCCTAGTCATCGTTACT-3'<br>3'-GTCTCTCAAGGGGGATCAGTAGCAATGAGCTTAAATCG-5' |
| (5,3) | 5'-CCCCTAGTCATCGTTACT-3'<br>3'-TGGCGTACACGGGGATCAGTAGCAATGAACCAATCTGC-5' |
| (4,5) | 5'-GCAAGTTTGG-3'<br>3'-AATACCTGTCCGTTCAAACCGCTTAAATCG-5' |
| (5,4) | 5'-GCAAGTTTGG-3'<br>3'-TGGCGTACACCGTTCAAACCTCCGTGTGCT-5' |

#The duplex segments proportional to the length of the edges are highlighted in red.

**Table 3.** DNA primer sequence design

| vertex | forward primer sequence (5' to 3') | reverse primer sequence (5' to 3') |
|---|---|---|
| 1 | CCGTGTCTACAACAGAAGGA | TCCTTCTGTTGTAGACACGG |
| 2 | CCCTCATTTGACGAGGAATG | CATTCCTCGTCAAATGAGGG |
| 3 | TGGTTAGACGCAGAGAGTTC | GAACTCTCTGCGTCTAACCA |
| 4 | AGGCACACGATTATGGACAG | CTGTCCATAATCGTGTGCCT |
| 5 | CGAATTTAGCACCGCATGTG | CACATGCGGTGCTAAATTCG |
| 6 | GGGGTCTTCAACTATTGTCC | GGACAATAGTTGAAGACCCC |

*Mixing controller* – After all DNA strands, ssDNA and dsDNA, as stated above, are prepared in individual eppendorf on *DNA solution bay*, a certain amount of solution, representing vertices and edges in the graph S, is aliquoted out and mixed in a new eppendorf to conduct the computing. For six vertex ssDNA sequences, which represent the vertices in graph S, 1μL of solution is taken from each stock solution. For the sixteen dsDNA sequences, which represent the edges, the amount of solution to be added is inversely proportional to their lengths as indicated in graph S. Such a strategy promotes the formation of the optimal path. The individual volume of dsDNA solution is detailed in Table 4. The resultant 22μL mixture is then subjected to a process known as hybridization, whereas the complementary ssDNA and DNA duplexes with sticky ends are readily annealed to each other through hydrogen bonding. To optimize the result of hybridization, 2.4μL of 10× ligation buffer is added into the mixture. The hybridization occurs instantly once all the DNA sequences are placed in *mixing controller*. Hence the incubation of *mixing controller* at room temperature takes only 15 minutes to ensure the completion of hybridization process.

**Table 4.** DNA solution pool (during hybridization)

| **Test Tube** | | **Solution Volume (μL)** |
|---|---|---|
| 1 to 6 | (ssDNA, 20mer) | 1.00 |
| 7 and 8 | (dsDNA, 46bp) | 0.78 |
| 9 and 10 | (dsDNA, 32bp) | 1.13 |
| 11 and 12 | (dsDNA, 42bp) | 0.86 |
| 13 and 14 | (dsDNA, 32bp) | 1.13 |
| 15 and 16 | (dsDNA, 38bp) | 0.95 |
| 17 and 18 | (dsDNA, 36bp) | 1.00 |
| 19 and 20 | (dsDNA, 38bp) | 0.95 |
| 21 and 22 | (dsDNA, 30bp) | 1.20 |

After the completion of hybridization process, DNA strands are held together through relative weak hydrogen bonds. In order to connect the optimal pathway into a one-piece DNA fragment as the template required by the next step, *PCR amplifier*, the adjacent DNA strands arranged by overhang annealing in the previous step are ligased by sequentially adding 2μL NEB kinase and 2μL T4 DNA liganse. In each step, hybridized DNA fragments are incubated with the enzyme for 20 minutes at 37°C.

In a DNA pool with as low as $50\ pmol$ solution, there are approximately $1.5 \times 10^{9}$ copies of associated DNA strands encoding each vertex and weighted edges, which is plenty enough to generate non-restricted random walks during hybridization among vertex and edge DNAs. Therefore, it is believed that all possible paths can be generated at the end of ligation process.

*Solution purifier* – In order to obtain the optimal outcome from the subsequent *PCR amplifier*, DNA solution from the previous stage is immediately subjected to *solution purifier*. In DNA-mediated computing, *solution purifier* is acting as a "noise signal remover" to isolate all the possible paths as dsDNA templates and ultimately to remove the impurities including incompletely hybridized oligonucleotides, various enzymes, and salt from desired solution. Contemporarily, there are several optimized purification kits available in the market, which are manufactured to satisfy the various requirements in accord with the range of DNA fragment binding-size. In this experiment, QIAquick® PCR purification kit is used for purification process.

After 2μL of final solution is taken for concentration analysis, the remaining solution of 26.4μL is pipetted into an empty test tube. A total volume of 132μL of PB buffer from QIAquick® PCR purification kit, or 5 times of the amount of the remaining solution, is added into the test tube. The well-mixed solution is poured into a QIAquick column and attached to a 2mL collection tube. Subsequently, the sample is centrifuged at 13,000 rpm (rounds per minute) for 60 seconds. The QIAquick column is placed back into a new test tube. The PE buffer of 0.75mL is added into the QIAquick column and centrifuged for another 60 seconds. Again, the flow through is discarded and the QIAquick column is placed back in the same tube. The QIAquick column is centrifuged in a 2mL collection tube for another 1 minute to remove the residual buffer. The QIAquick column is then placed in a new 1.52mL micro-centrifuge tube. Finally, 50μL EB buffer is added into the membrane at the center of the column, retained at room temperature for another 1 minute, and followed by one additional centrifugal process to elute the DNA solution. The final elute contains roughly 40 to 50μL of purified dsDNA duplex.

*PCR amplifier* – PCR is a frequently employed laboratory technique in DNA-mediated computing to amplify specific "signal" – DNA template, by varying the "input signal" – various synthesized DNA primers, which is specified by the user of optimal route planning processor. To demonstrate the DNA-mediated processor for optimal path question in GPS devices, two cases are studied here. It is desired to determine the optimal path traveling from two distinct locations, home (vertex 1) and company (vertex 6), to the same destination – hospital (vertex 4). In the first case, the optimal path starts from home (vertex 1) and terminates at hospital (vertex 4). The selected primers for *PCR amplifier* are the forward primer of vertex 1 and the reverse primer of vertex 4, as specified in Table 3. Similarly, in the second case, the optimal path starts from company (vertex 6) and terminates at hospital (vertex 4). The forward primer of vertex 6 and the reverse primer of vertex 4 are used.

The selected primer and purified DNA solution from the previous stage are mixed together with the reagents of QIAGEN fast cycling PCR kit in eppendorf. Each PCR sample contains 14μL PCR reagents (10μL of PCR master mix and 4μL of Q solution), 2μL of the corresponding forward primer, 2μL of the corresponding reverse primer, and 2μL of purified DNA solution from the previous stage, resulting in a mixture of 20μL prior to PCR.

The detailed PCR protocol is described as follows:

*Step 1*: Increase solution temperature to 95°C and maintain for 5 minutes.

*Step 2*: Increase solution temperature to 96°C and maintain for 5 seconds.

*Step 3*: Gradually reduce solution temperature to 48°C and maintain for 5 seconds.

*Step 4*: Increase solution temperature to 68°C and maintain for 5 seconds.

*Step 5*: Repeat Step 2 to Step 4 for 29 cycles.

*Step 6*: Increase solution temperature to 72°C and maintain for 1 minute.

*Step 7*: Gradually reduce solution temperature to room temperature 25°C.

Based upon the different primers selected for these two cases above, the PCR samples are labelled with "1 → 4" and "6 → 4", respectively. As the result of PCR, for " 1 → 4", only the paths which starts with vertex 1 and terminates at vertex 4 are amplified. The same condition applies to “6 → 4”. After *PCR amplifier*, it is recommended to repeat the purification process to remove any possible impurities. However, the purification process is not compulsory.

*Gel electrophoresis* – *Gel electrophoresis* employs native polyacrylamide gel as sieving material to distribute the linear DNA molecules with respect to their lengths. In this experiment, 12% gel is prepared according to the recipe as specified in Table 5. After the completion of electrophoresis, the gel is stained in the TBE buffer containing 3× GelRed for 30 minutes. Subsequently, the results are observed in a 2-dimentional multi-fluorescence scanner Typhoon 9410. Under the green-excited mode, the DNA bands in gel slab have a strong emission subjected to excitation at 532nm wavelength UV (ultraviolent) light. The 2D image of 12% native acrylamide gel of four lanes is displayed in Figure 4(a).

**Table 5.** 12% native polyacrylamide gel recipe

| **Ingredient** | **Volume (μL)** |
|---|---|
| $H_2O$ | 15.7 |
| 30% acrylamide | 16 |
| 5 × TBE buffer | 8 |
| 10% ammonium persulfate | 0.28 |
| TEMED | 0.026 |

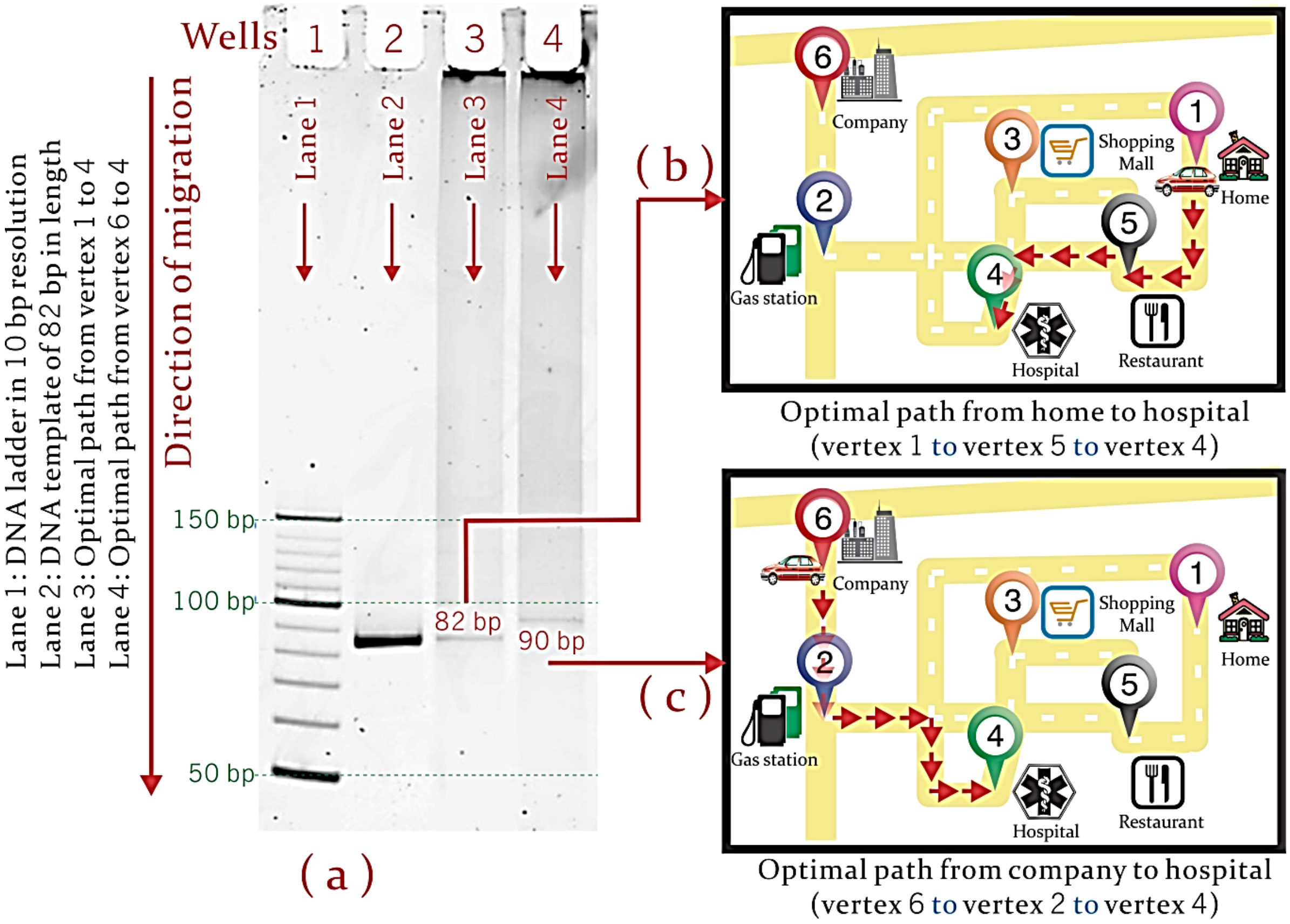

**Figure 4.** The optimal paths from home or company to hospital on 12% PAGE (a: *lane 3* and *lane 4*) and illustrative map (b and c). *Lane 1*: DNA ladder; *Lane 2*: 82bp DNA.

## ■ RESULTS

Figure 4 shows the PCR outcome of the optimal path from either home (vertex 1) or company (vertex 6) to hospital (vertex 4) selected by the DNA-mediated processor. Four wells are built in gel slab to contain different solution – *well 1*: DNA ladders in 10bp interval; *well 2*: dsDNA template of 82bp; *well 3*: PCR product of the sample "1 → 4"; *well 4*: PCR product of the sample "6 → 4". *Well 1* is used to display DNA ladders, ranging from 50bp to 150bp, in 10bp interval. According to its working principle, the migration distance of DNA solution toward anode electrode is inversely proportional to the length of linear dsDNA strands. DNA ladders in *well 1* can be used to gauge the length of PCR outcome from the two selections in *lane 3* and *lane 4*. For the same purpose, *well 2* contains the dsDNA duplex of a specified length of 82bp.

The gel slab immersed in TBE buffer is subjected to gel electrophoresis process by setting the power of 18W with variable voltage and current at 560V and 32mA, respectively. The entire electrophoresis process takes roughly one and a half hours. After the gel electrophoresis process, it is possible to determine the total length of optimal path by comparing the migration lengths of the most prominent band in *lane 3* and *lane 4* with respect to the bands in reference lanes, which are *lane 1* and *lane 2*. The maximum migration length toward the anode electrode of the most prominent band along *lane 3* or *lane 4* is the dsDNA duplex of the shortest length. Therefore, the DNA-mediated processor provides the solution to the optimal path problem for both user inputs discussed above.

## ■ DISCUSSION

In both cases of "1 → 4" and "6 → 4" (*lane 3* and *lane 4*, respectively), only one prominent band is observed to indicate the most optimal path selected by the GPS. The length of optimal path obtained from the DNA processor is gauged by the DNA ladders in *lane 1* ranging from 50bp to 150bp in the interval of 10bp, and/or a dsDNA duplex in *lane 2* of a specified length 82bp, as the expected shortest distance from home (vertex 1) to hospital (vertex 4), that is, the path "1 → 5 → 4". Such length and sequence of the reference DNA are determined by using the encoding scheme together with Figure 1. The reference dsDNA duplex in *lane 2* is used to determine the migration length of the most prominent band *in lane 3*. In simple words, it is going to testify the following hypothesis: The resultant DNA from optimal route planning processor, which represents the shortest path from home to hospital as shown in *lane 3*, has the same migration distance as predicted in *lane 2*.

*Lane 3* is used to migrate the PCR product, which is selected by the primers representing the starting and terminating points as vertex 1 and vertex 4, respectively. As observed in Figure 4(a), a uniquely visible band in *lane 3* has the same migration distance as that in *lane 2*. From the entire template pool of all possible routes between any two desired locations, the pair of input primers is able to amplify specifically one template of 82bp as the only output signal from the DNA processor. Therefore, it is believed the programmable DNA-mediated optimal route planning processor has "selected" the expected path, which is the path "1 → 5 → 4", as shown in Figure 4(b).

Analogously, a unique band in *lane 4* can be observed by using the forward primer of vertex 6 and the reverse primer of vertex 4. As compared with the DNA ladder in *lane 1*, the migration distance of band in *lane 4* is equivalent to that of dsDNA duplex of 90bp in *lane 1*. As depicted in Figure 1, it is possible to determine that the optimal path from company (vertex 6) to hospital (vertex 4) is likely to be the path "6 → 2 → 4" as shown in Figure 4(c), which is 90bp based upon the DNA encoder listed in Tables 1 and 2.

In the two cases studied here, all of the possible paths between any of two locations are generated in the DNA-encoded processor simultaneously. Once that all the vertex and edge DNA strands are hybridized in the step of *mixing controller*, the readout of the outcome, *the* *optimal path*, is unambiguous if the specific pair of PCR primers is used as an input command from the user.

## ■ CONCLUSIONS

The novel programmable DNA-mediated processor is developed to encode a complete bi-directional road map including six locations. The optimal route between any of two locations can be simultaneously selected by the programmable DNA-mediated processor in parallel and selectively revealed by the conventional biochemistry assays, *PCR amplifier* and *gel electrophoresis*. This programmable DNA-mediated processor has several advantages over the existing silicon-mediated methods in solving similar problems, such as conducting massive data storage and simultaneous processing for multiple path selections. In theory, it requires much fewer materials and much lower space than conventional silicon devices. The utilizing length of DNA strands, as the biasing parameter, is believed to be superior to the other alternatives, such as varying the pH value or concentration of DNA solution. Furthermore, with the proper encoding of the map into DNA edge duplexes, the optimal path can be directly deduced by the length of PCR outcome. Hence the current processor avoids the tremendous processing cost and time in DNA sequencing, which is required by many precedent DNA-based systems to address mathematical problems. It is worth to mention to this end that the DNA GPS system proposed in this paper may provide a new way to understand the working mechanism of a brain's GPS system due to the Nobel Prize winning discovery of the place[14] and grid[15] cells.

## AUTHOR INFORMATION

**Corresponding Author**

*(J.-J.S.) E-mail: mjjshu@ntu.edu.sg.

**Notes**

The authors declare no competing financial interest.

REFERENCES

(1) Moore, G. E. Cramming More Components onto Integrated Circuits. *Electronics* **1965**, *38*, 114–117.

(2) Kuhnert, L.; Agladze, K. I.; Krinsky, V. I. Image-Processing Using Light-Sensitive Chemical Waves. *Nature* **1989**, *337*, 244–247.

(3) Divincenzo, D. P. Quantum Computation. *Science* **1995**, *270*, 255–261.

(4) Brooks, R. Artificial Life - From Robot Dreams to Reality. *Nature* **2000**, *406*, 945–947.

(5) Seeman, N. C. DNA in a Material World. *Nature* **2003**, *421*, 427–431.

(6) Yurke, B.; Turberfield, A. J.; Mills, A. P.; Simmel, F. C.; Neumann, J. L. A DNA-Fuelled Molecular Machine Made of DNA. *Nature* **2000**, *406*, 605–608.

(7) Church, G. M.; Gao, Y.; Kosuri, S. Next-Generation Digital Information Storage in DNA. *Science* **2012**, *337*, 1628–1628.

(8) Benenson, Y.; Gil, B.; Ben-Dor, U.; Adar, R.; Shapiro, E. An Autonomous Molecular Computer for Logical Control of Gene Expression. *Nature* **2004**, *429*, 423–429.

(9) Reif, J. H. Scaling up DNA Computation. *Science* **2011**, *332*, 1156–1157.

(10) Adleman, L. M. Molecular Computation of Solutions to Combinatorial Problems. *Science* **1994**, *266*, 1021–1024.

(11) Ouyang, Q.; Kaplan, P. D.; Liu, S. M.; Libchaber, A. DNA Solution of the Maximal Clique Problem. *Science* **1997**, *278*, 446–449.

(12) Shu, J.-J.; Wang, Q.-W.; Yong, K.-Y. DNA-Based Computing of Strategic Assignment Problems. *Phys. Rev. Lett.* **2011**, *106*, 188702.

(13) Tanaka, F.; Kameda, A.; Yamamoto, M.; Ohuchi, A. Design of Nucleic Acid Sequences for DNA Computing Based on a Thermodynamic Approach. *Nucleic Acids Res.* **2005**, *33*, 903–911.

(14) O'Keefe, J.; Dostrovs, J. The Hippocampus as a Spatial Map. Preliminary Evidence from Unit Activity in the Freely-Moving Rat. *Brain Res.* **1971**, *34*, 171–175.

(15) Hafting, T.; Fyhn, M.; Molden, S.; Moser, M.-B.; Moser, E. I. Microstructure of a Spatial Map in the Entorhinal Cortex. *Nature* **2005**, *436*, 801–806.

Table of Contents (TOC) Graphic

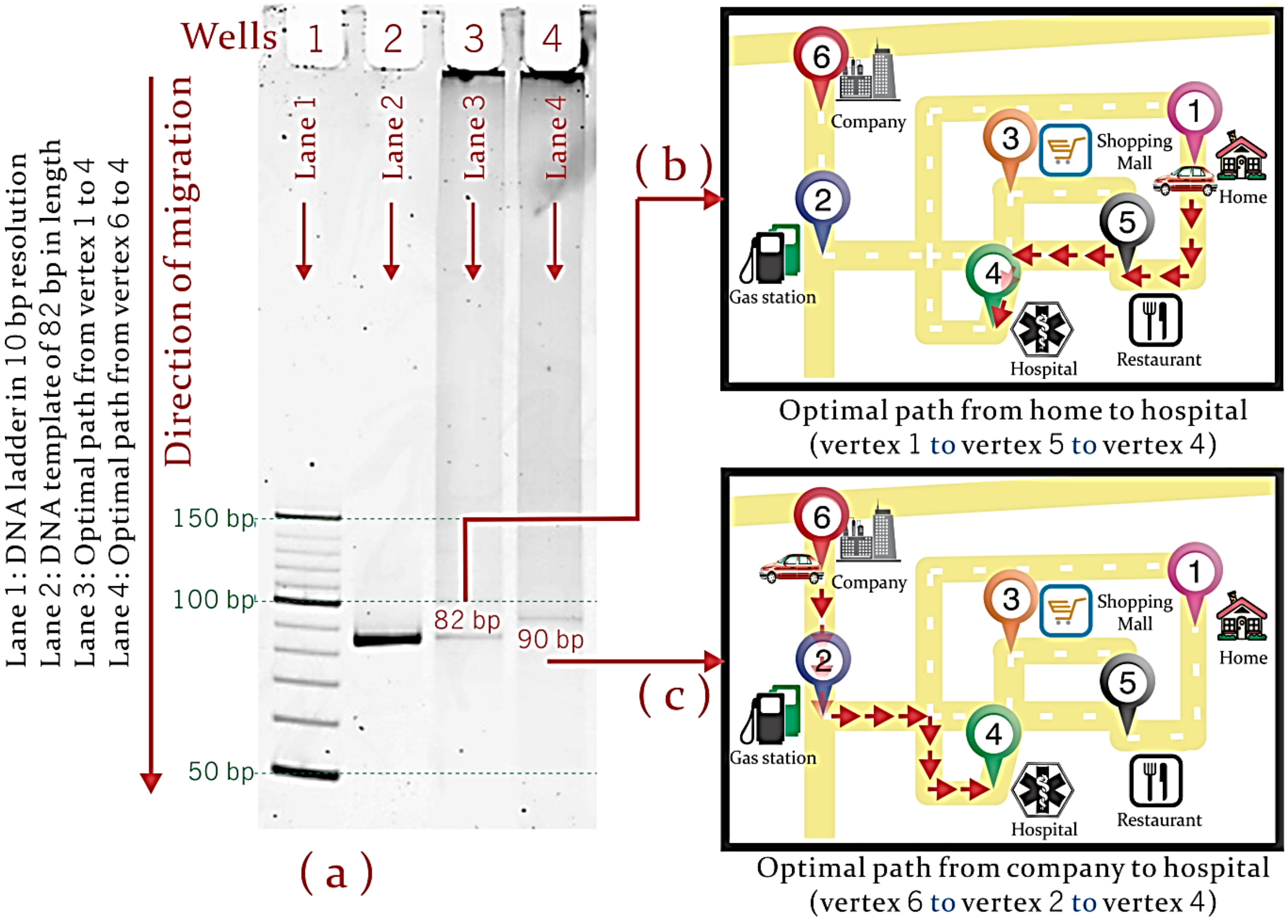